\documentstyle[onecolumn,subfigure]{mn}
\begin{document}
\newcommand{\mincir}{\raise
-2.truept\hbox{\rlap{\hbox{$\sim$}}\raise5.truept
\hbox{$<$}\ }}
\newcommand{\magcir}{\raise
-2.truept\hbox{\rlap{\hbox{$\sim$}}\raise5.truept
\hbox{$>$}\ }}
\newcommand{\minmag}{\raise-2.truept\hbox{\rlap{\hbox{$<$}}\raise
6.truept\hbox
{$>$}\ }}
\newcommand{\be}{\begin{equation}}
\newcommand{\ee}{\end{equation}}
\newcommand{\ba}{\begin{eqnarray}}
\newcommand{\ea}{\end{eqnarray}}
\newcommand{\brr}{\begin{array}}
\newcommand{\nn}{\nonumber \\} 
\newcommand{\err}{\end{array}}
\newcommand{\bc}{\begin{center}}
\newcommand{\ec}{\end{center}}
\newcommand{\br}{\mbox{\bf r}}
\newcommand{\bv}{\mbox{\bf v}}
\newcommand{\bs}{\mbox{\bf s}}
\newcommand{\bq}{\mbox{\bf q}}
\newcommand{\bx}{\mbox{\bf x}}
\newcommand{\by}{\mbox{\bf y}}
\newcommand{\bk}{\mbox{\bf k}}
\newcommand{\tR}{\mbox{\tiny R}}
\newcommand{\tM}{\mbox{\tiny M}}
\newcommand{\tN}{\mbox{\tiny N}}
\newcommand{\tL}{\mbox{\tiny L}}
\newcommand{\lb}{{\left<\right.}}
\newcommand{\rb}{{\left.\right>}}
\newcommand{\hm}{\,h^{-1}{\rm Mpc}}
\newcommand{\vel}{\,{\rm km\,s^{-1}}}

\renewcommand{\baselinestretch}{1.0}
\input{psfig.sty} 
\title[Measuring the Redshift Evolution of Clustering: the HDF-South]
{Measuring the Redshift Evolution of Clustering: the Hubble Deep Field
South \footnote{Based on observations with the NASA/ESA Hubble Space
Telescope, and on observations collected with the ESO-VLT as part of
the programme 164.O-0612}}

\author[Arnouts et al.]  { S. Arnouts$^{1}$, L. Moscardini$^{2}$,
E. Vanzella$^{1,2}$,  S. Colombi$^{3}$,  S. Cristiani$^{4,2}$, \newauthor
A. Fontana$^5$, E. Giallongo$^5$,  S. Matarrese$^{6}$ and P. Saracco$^7$ \\
$^1$ ESO -- European Southern Observatory, Karl-Schwarzschild-Str. 2, 
D--85748 Garching bei M\"unchen, Germany \\
$^2$ Dipartimento di Astronomia, Universit\`a di Padova,
vicolo dell'Osservatorio 2, I--35122 Padova, Italy \\
$^3$ Institut d'Astrophysique de Paris, 98bis bd Arago, 75014 Paris, France\\
$^4$ Space Telescope European Coordinating Facility, Karl-Schwarzschild-Str. 2,
 D--85748 Garching bei M\"unchen, Germany \\
$^5$ Osservatorio Astronomico di Roma, via dell'Osservatorio, I-00040
Monteporzio (RM), Italy \\
$^6$ Dipartimento di Fisica G. Galilei,  Universit\`{a} di 
Padova, via Marzolo 8, I--35131 Padova, Italy \\
$^7$ Osservatorio Astronomico di Brera, via Bianchi 46,  I--23807 
Merate (LC), Italy\\
 }

\date{Accepted 2001 september 25. Received 2001 may 10;
in original form 2001 may 10}

\maketitle

\begin{abstract}
We present an analysis of the evolution of galaxy clustering in the
redshift interval $0 \le z \le 4.5$ in the HDF-South. The HST optical
data are combined with infrared ISAAC/VLT observations, and
photometric redshifts are used for all the galaxies brighter than
$I_{AB}\le 27.5$.  The clustering signal is obtained in different
redshift bins using two different approaches: a standard one, which
uses the best redshift estimate of each object, and a second one,
which takes into account the redshift probability function of each
object. This second method makes it possible to improve the
information in the redshift intervals where contamination from objects
with insecure redshifts is important.  With both methods, we find that
the clustering strength up to $z\simeq 3.5$ in the HDF-South is
consistent with the previous results in the HDF-North. While at
redshift lower than $z\sim 1$ the HDF galaxy population is
un/anti-biased ($b\le 1$) with respect to the underlying dark matter,
at high redshift the bias increases up to $b(z\sim 3)\simeq 2-3$,
depending on the cosmological model.  These results support previous
claims that, at high redshift, galaxies are preferentially located in
massive haloes, as predicted by the biased galaxy formation scenario.
In order to quantify the impact of cosmic errors on our analyses, we
have used analytical expressions from Bernstein (1994).  Once the
behaviour of higher-order moments is assumed, our results show that
errors in the clustering measurements in the HDF surveys are indeed
dominated by pure shot-noise in most regimes, as assumed in our
analysis.  We also show that future observations with instruments like
the Advanced Camera on HST will improve the signal-to-noise ratio by
at least a factor of two; as a consequence, more detailed analyses of
the errors will be required. In fact, pure shot-noise will give a
smaller contribution with respect to other sources of errors, such as
finite volume effects or non-Poissonian discreteness effects.

\end{abstract}
\begin{keywords}
cosmology: observations -- photometric redshifts -- large--scale
structure of Universe -- cosmic errors -- galaxies: formation --
evolution -- haloes
\end{keywords}

\section{Introduction}
It is well known that the evolution of the dark matter clustering can
be reliably used to put strong constraints on cosmological models. In
fact the growth of density fluctuations depends on the main
cosmological parameters, namely the contribution of matter and
cosmological constant to the present total energy density
($\Omega_{\rm 0m}$ and $\Omega_{\rm 0\Lambda}$, respectively). This
result, confirmed by high-resolution N-body simulations (e.g. Jenkins
et al. 1998), has been used to build a semi-empirical model which
suitably relates the linear perturbation scale to the final non-linear
scale of the same perturbation after collapse (Hamilton et
al. 1991). This technique can be used to compute analytically the
evolved correlation function starting from a given primordial density
power-spectrum (e.g. Peacock \& Dodds 1994, 1996; Jain, Mo \& White
1995).

However, the application of this idea to real data is greatly
complicated by the fact that the observed objects (galaxies, quasars,
clusters, etc.) are not direct tracers of the dark matter
distribution. Usually, the ignorance about the relation between the
object density, $\delta_{\rm o}$, and the dark matter one,
$\delta_{\rm m}$, is parametrized introducing the so-called bias
parameter $b$, for which a simple linear relation is a common
assumption: $b=\delta_{\rm o}/\delta_{\rm m}$ (Kaiser 1984). Note that
this relation includes the details of structure formation and, as a
consequence, is quite uncertain.

A possible shortcut to the solution of this problem is to relate the
value of $b$ to some intrinsic property. For example, analytical
models (e.g.  Mo \& White 1996; Catelan et al. 1998; Jing 1999; Sheth
\& Tormen 1999; Sheth, Mo \& Tormen 2001), confirmed by the results of
N-body simulations, suggest that the bias factor of dark matter haloes
is a function only of their mass and formation redshift (apart from
the cosmological parameters). If there is a way to relate a typical
observational quantity of the considered objects (such as flux or
luminosity) directly to the mass of their hosting dark matter haloes,
the study of the clustering evolution fully recovers its ability to
discriminate between different cosmological models.  For instance, in
the case of galaxy clusters detected in the X-ray band, the flux at a
given redshift corresponds to a given halo mass, under the assumptions
of virial isothermal gas distribution and spherical collapse.
Hydrodynamical simulations confirm the resulting relations between
mass and luminosity or temperature, even if with a large scatter. Then
the comparison of the observed cluster two-point correlation function
to theoretical predictions can be used to put some constraints to the
cosmological parameters.  For example, Moscardini et al. (2000a,b)
find that the clustering properties of the clusters observed in
different samples (RASS1 Bright Sample, XBACs, BCS and REFLEX) favour
cosmological models with a low value of $\Omega_{\rm 0m}$.

In the case of galaxies, applying a similar technique is much more
difficult, mainly because the relation between mass and luminosity is
not one-to-one. Moreover, it is not clear how many galaxies can occupy
a single halo of a given mass. However, once a cosmological framework
is fixed, the study of the clustering evolution of galaxies can be
used to obtain more information about the nature of these objects. For
example, it is possible to estimate a typical value for the mass of
the dark matter haloes hosting the galaxies. Moreover the clustering
data can be used to discuss if the merging process is important at
various redshifts or if the galaxy number tends to be conserved during
the evolution. In fact, these two opposite models predict a completely
different redshift evolution of the bias factor (see e.g. Matarrese et
al. 1997 and Moscardini et al. 1998).

 From the point of view of the observational data required for this
 kind of study, enormous progress has been made in recent years.
 Large spectroscopic surveys gave an accurate description of the
 spatial distribution of the galaxies in the local
 Universe. Statistical analyses have shown that the correlation length
 depends on morphological type and/or absolute magnitude: more
 luminous and/or early-type galaxies appear to have higher clustering
 than faint and/or late-type galaxies (Santiago \& da Costa 1990;
 Loveday et al. 1995; Benoist et al. 1996; Norberg et al., 2001).
 However, these local observations can be reasonably well reproduced
 by a large variety of sensible cosmological models, while possible
 differences are expected at higher redshifts, as previously
 discussed. This has been one of the main reasons which motivated the
 extension of spectroscopic surveys to high redshifts. Nowadays,
 different samples are available to estimate the clustering properties
 of galaxies up to $z\approx 1$ [Canada-France Redshift survey (Le
 F\`evre et al. 1996); Hawaii K survey (Carlberg et al. 1997); Norris
 Redshift Survey (Small et al. 1999); Caltech Faint Redshift survey
 (Hogg, Cohen \& Blandford 2000); Canadian Network for Observational
 Cosmology field galaxy redshift survey (Carlberg et al. 2000)]. Even
 if the sampled regions are relatively small, the results are in good
 agreement in showing a decline of the correlation length with
 redshift.

Up to a few years ago, clustering studies at higher redshifts were
limited to peculiar objects like radiogalaxies or quasars. The
discovery of reliable colour techniques (U-dropouts) made it possible
to identify a large sample of `normal' galaxies at $z\approx 3$, the
so-called Lyman-Break galaxies (LBGs). By measuring the correlation
function or computing the count-in cell statistics, different works
(Adelberger et al. 1998; Giavalisco et al. 1998; Giavalisco \&
Dickinson 2000) showed that LBGs have a correlation length at least
comparable with that of present-day spiral galaxies. This result
corresponds to quite a high  value for the bias factor at $z\sim 3$,
suggesting that their formation occurs in massive dark-matter haloes.

An alternative way to probe larger volumes and/or fainter galaxy
populations makes use of the photometric redshift technique (e.g.
Lanzetta, Yahil \& Fern\'andez-Soto 1996; Sawicki, Lin \& Yee 1997;
Arnouts et al. 1999, hereafter A99; Bolzonella, Miralles \& Pell\`o
2000).  This method, based on the comparison of theoretical and/or
real spectra with the observed galaxy colours in different bands,
makes it possible to estimate their redshifts at higher magnitudes
than those reached spectroscopically by the largest available
telescopes.  This is done in a probabilistic way; as a consequence,
the estimates are affected by errors, which typically have been found
to increase with redshift. Note that to date, these inherent
uncertainties in the redshift estimates were completely ignored or
estimated via simulations and used as an {\it a posteriori} global
correction to the correlation measurements (A99).  A more correct
approach would require the estimate of the redshift uncertainty for
each object and the inclusion of this information in the computation
of the correlation function, as discussed in this paper.

Thanks to the application of the technique of photometric redshifts to
its very deep observations, the Hubble Deep Field (HDF) North
(Williams et al. 1996) has become a test case for the evolution of
the galaxy distribution.  The data of more than one thousand objects
down to $I_{AB}\approx 28.5$ have been used to study the redshift
evolution of the clustering up to $z\sim 4.5$ (A99; Magliocchetti \&
Maddox 1999; Roukema et al. 1999; see also the analysis made by
Connolly, Szalay \& Brunner 1998 up to $z\sim 1.2$). The results show
that the comoving correlation length, after a small decrease in the
interval $0 \mincir z \mincir 1$, increases up to $z\sim 4$. It is
worthwhile to stress that the term ``evolution'' has not to be taken
literally.  Given a survey defined by its characteristic limiting
magnitude and surface brightness, the galaxies observed at high $z$
typically have higher luminosities.  Therefore, the intrinsic
differences of the galaxy properties at different $z$ can mimic an
evolution, i.e. the evolution measured in a flux-limited survey is not
only due to the evolution of a unique population but can be due to a
change of the considered population. In fact the theoretical modelling
of the HDF galaxies shows that to reproduce their clustering
properties at different redshifts, the mean mass of dark matter haloes
hosting the galaxies is required to increase with $z$ (A99).

The reliability of the previous results, however, can be affected by
the smallness of the observed field. In particular, it is not clear to
what extent a region of a few square arcminutes can be considered
representative of the properties of the whole Universe. The data more
recently obtained in the HDF-South (Casertano 2000) offer a unique
opportunity to test the robustness of HDF-North results, due to their
mutual independence.  For example, the field-to-field variations can
be used to estimate the size of the cosmic variance on these scales.
The main goal of this paper is to study in detail the clustering
properties of HDF-South and to compare them with those obtained for
the northern field to confirm or disprove the general picture
described above.

The paper is organized as follows: In Section 2 we present the
photometric database used in this analysis and briefly describe the
photometric redshift technique. In Section 3 we introduce the two
methods used to estimate the angular correlation function: the
standard approach and an alternative method taking into account the
photometric redshift uncertainties. Still in Section 3 we present the
results of this analysis and estimate the bias factor.  Section 4 is
devoted to a theoretical discussion of the cosmic errors in the
clustering estimates in Hubble Deep Fields. Conclusions are presented
in Section 6.
 
\section{The catalogue and photometric redshifts}

\subsection{The data}
 
Deep high-resolution optical dataset ($F_{300}$, $F_{450}$, $F_{606}$
and $F_{814}$) from HST and deep infrared observations have been
combined. The IR observations have been carried out in $J_s,H,K_s$
passbands with the ISAAC instrument on the VLT (UT2) during the period
July-September 1999. The total integration times are 7h, 6h and 8h in
$J$, $H$ and $K_s$, respectively.  The final coadded images have a
seeing of 0.6 arcsec in $J_s$,$H$,$K_s$. The Vega magnitude limits in
$2 FWHMs$ at 5$\sigma$ level are 24,23,22.5 in $J_s$,$H$,$K_s$,
respectively (Saracco et al. 2001). 

The photometric catalogue containing the optical and infrared colours
is described in detail in Vanzella et al. (2001). We recall here that
the detections are based on the summed $V+I$ images and the deblending
process has been tuned and optimized in order to obtain a photometric
catalogue particularly reliable for photometric redshifts.  Indeed a
modified version of the SExtractor software (Bertin \& Arnouts 1996)
has been applied to optimize the SExtractor parameters (namely
{\it deblend-mincont}, {\it detect-minarea}) in different regions of the
frame. This procedure allows to improve the deblending of close pairs
as well as to keep in single units large spiral galaxies and affects
only the very small angular scales ($\theta \le 3 arcsec$). A catalogue of
1474 sources has been extracted up to $I_{AB}\simeq 28.5$.

\subsection{The photometric redshift measurement}

\begin{figure*}
\centerline{\psfig{figure=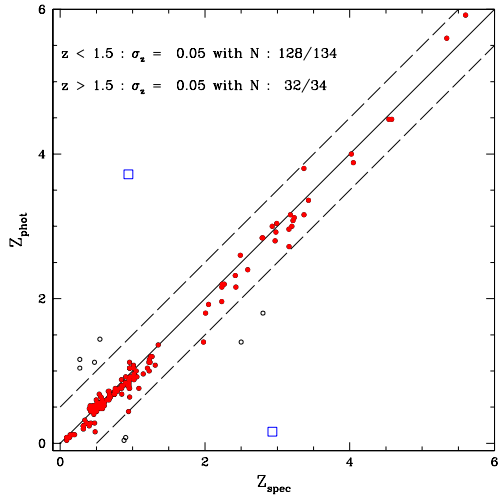,angle=0,width=7.cm}}
\caption[]{ Dispersion between the spectroscopic redshifts and
     photometric estimates in the HDF-North (146 spectra) and
     HDF-South (24 spectra) (see text).  The redshift dispersion
     ($\sigma_z$) is obtained by using a $3\sigma$-clipping rejection
     for two samples: z$\le 1.5$ and $z\ge 1.5$. Catastrophic
     redshifts (represented by square symbols) have not been used in
     the measurement.  Rejected objects during the $\sigma$-clipping
     are shown with open circle symbols.  The solid line corresponds
     to $\Delta z= 0$ and the long-dashed lines to $\Delta z= 0.5$. }
\label{fdz}
\end{figure*}

The technique of photometric redshifts adopted in this paper has been
described in more detail in A99.  The technique is based on $\chi^2$
minimisation which compares the observed magnitudes to the GISSEL96
synthetic library (Bruzual \& Charlot 1993). In order to quantify the
redshift uncertainties, in Figure\ref{fdz} we compare, for galaxies
accessible to spectroscopy, the spectroscopic redshifts and those
obtained using the photometric technique. The HDF-North sample is
based on the list of Cohen et al. (2000) which is composed of 146
spectra. The HDF-South sample is based on 22 spectra from the list of
Cristiani et al. (1999) observed with the VLT telescope and from
Dennefeld et al.  (2001) observed with NTT telescope. We also add 2
spectra observed with the Anglo Australian Telescope (Glazebrook et
al., 1998). In the area of WFPC2 the HDF-South sample consists of 24
spectra, two of which are at $z_{\rm spec}>1.5$. The redshift
accuracy is defined as in Fern\'andez-Soto, Lanzetta \& Yahil (1999):
$(z_{\rm spec}-z_{\rm phot})/(1+z_{\rm spec})$, from which we extract
the mean ($\Delta z$) and the dispersion ($\sigma_z$) by using a
$\sigma$-clipping algorithm at $3\sigma$ rejection level.  We obtain
$\sigma_z=0.05$ and $\Delta z=0.03$ for $z_{\rm spec}\le 1.5$ and
$\sigma_z=0.05$ and $\Delta z=0.02$ for $z_{\rm spec}\ge 1.5$.  Two
catastrophic redshifts were initially rejected from the statistics
(shown by large open squares in Figure~\ref{fdz}) and six objects at
$z\le 1.5$ and two objects at $z\ge 1.5$ were rejected during the
$\sigma$-clipping process (open circles in Figure~\ref{fdz}).  The
total number of rejected objects is 10/170, corresponding to 6 per
cent.

%--------------------------------------------------------
%--------------------------------------------------------

\begin{figure*}
\centerline{ \psfig{figure=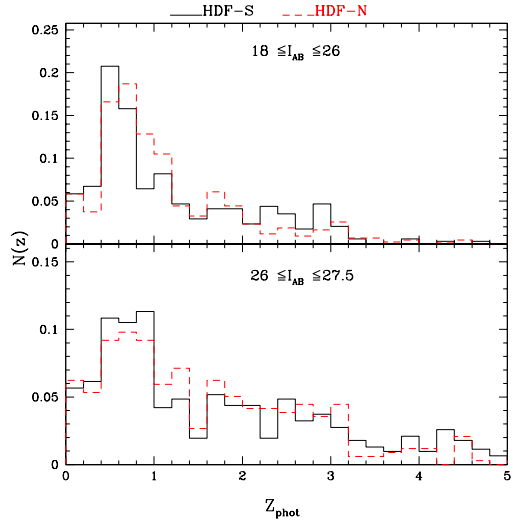,angle=0,width=7.cm}}
\caption[]{ Comparison of the redshift distributions of the HDF-North
(dashed line) and HDF-South (solid line) for galaxies brighter than
$I_{AB}\le 26$ (upper panel) and galaxies with magnitudes in the range
$26\le I_{AB} \le 27.5$ (lower panel). }
\label{fNzns}
\end{figure*}

In Figure~\ref{fNzns} we compare the redshift distributions obtained
for the HDF-North and HDF-South for two intervals of magnitude,
$I_{AB}\le 26$ and $26 \le I_{AB}\le 27.5$ (upper and lower panels,
respectively).  The two redshift distributions are similar.  The
Kolmogorov-Smirnov (KS) two-tail statistics does not reject the null
hypothesis that the redshift distributions in the HDF-North and South
are drawn from the same parent population.  The KS-probability of the
null hypothesis turns out to be 0.12 and 0.20 for the samples with
$I_{AB}\le 26$ and $26\le I_{AB}\le 27.5$ respectively.  The median
redshift in the HDF-North seems to be slightly higher for the bright
sample, which is not surprising due to the presence of large-scale
structures at $z\sim 1$ in the HDF-North (Cohen et al., 2000), also
evidenced by systematic color differences (Vanzella et al., 2001).

%--------------------------------------------------------
%--------------------------------------------------------
 
\section{The angular correlation function}    
\subsection{Selection of the sample}

To compute the angular correlation function (ACF), we have limited our
analysis to the region of the HDF-South with the highest
signal-to-noise, excluding the area of the PC and the outer part of
the three WFPC.  The details of how the HDF-North and HDF-South
photometric catalogues (Fern\'andez-Soto et al. 1999; Vanzella et
al. 2001) have been constructed are slightly different and the scale
of the total $I$ magnitudes may present systematic differences. This
is especially true at faint magnitudes. For this reason, rather than
applying formally equal magnitude limits to the two samples, we prefer
to adopt for HDF-South a limit that defines a roughly equal number of
sources as in HDF-North and a comparable number of objects in each
redshift interval (at least 100 except for the range $3.5\le z \le
4.5$).  This can be accomplished by selecting in the HDF-South catalogue
all galaxies brighter than $I_{AB}\simeq 27.5$. The total number of
objects is 844 in an effective area of 4.45 $arcmin^2$.  The nominal
magnitude limit used in the HDF-North by A99 is $I_{AB}\simeq 28.5$
which provides 926 objects.  Beyond $I_{AB}\sim 26$ the source counts
in the HDF-S photometric catalog adopted in the present work are
systematically higher than the corresponding counts in the HDF-North
catalog used in A99 by a factor $\sim 1.5$. The discrepancy is to be
ascribed to differences in the approach used to carry out the
photometry in the two cases (Vanzella et al., 2001).  The redshift
bins used are the same as those adopted in the analysis of A99.

\subsection{Classical ACF computation}
\label{s:cacf}
The angular correlation function $\omega(\theta)$ is related to the
excess of galaxy pairs in two solid angles separated by the angle
$\theta$ with respect to a random distribution.  The angular
separation used for the computation of $\omega(\theta)$ covers the
range from 3 arcsec up to 80 arcsec. We use logarithmic bins with
steps of $\Delta \log (\theta) = 0.3$.  The lower limit of 3 arcsec is
a conservative estimate of the scale over which we are confident about
the deblending approach for resolved bright spirals or faint galaxy
``groups''.  The upper cut-off corresponds to almost half the size of
the HDF regions and to the maximum separation where the ACF provides a
reliable signal.

To derive the ACF in each redshift interval, we used the estimator
defined by Landy \& Szalay (1993, hereafter LS93):
\begin{equation}
 \omega_{\rm est}(\theta) \ = \ A_1 \ \frac{DD(\theta)}{RR(\theta)} 
 - 2 A_2 \ \frac{DR(\theta)}{ RR(\theta) }  + 1\ ,
\label{ew}
\end{equation}    
where DD is the number of different galaxy pairs, DR is the number of
galaxy-random pairs and RR refers to random-random pairs with
separation between $\theta$ and $\theta + \Delta \theta$.  The
normalisation factors $A_1$ and $A_2$ are given by
\begin{equation}
A_1=\frac{N_r (N_r-1)}{N_g (N_g-1)} \hbox{\ \ \ ;\ \ \ }
A_2=\frac{N_r-1}{2\ N_g} \ ,
\end{equation}
where $N_g$ and $N_r$ are the total number of objects in the data and
random catalogues, respectively. In the present work the random
catalogues contain $N_r=20000$ sources covering the same area as our
HDF sample.
 
In the weak clustering limit, the above estimator has a nearly
Poissonian variance (see LS93), so the uncertainty is estimated as:
\begin{equation}
  d\omega_{\rm est}(\theta) = \sqrt{\frac{1 + \omega_{\rm
  est}(\theta)}{<RR(\theta)>}}\hbox{\ \ \ ;\ \ \ } <RR(\theta)> =
  RR(\theta)/A_1
\label{ewterr}
\end{equation}

The results of our analysis will be discussed in Section 3.4, where
they will be compared with those obtained by the alternative approach,
described in the next subsection.

\subsection{Alternative ACF method}

\subsubsection{Redshift probability distribution function}
 
In our previous analysis of HDF-North (A99), we used Monte Carlo
simulations to discuss the effects of uncertainties in the $z_{\rm
phot}$ estimates on the clustering results. In particular we used the
simulations to obtain the statistical errors in each redshift
interval according to the limiting magnitude and to define an upper
limit to the amplitude of the ACF assuming that the contamination
effects are due to an uncorrelated population.
In the present work, we define an alternative method which includes
directly in the ACF measurement the redshift probability distribution
of each object.
For each object we measure a redshift probability distribution
function (hereafter $PDFz$) estimated as follows:

\be PDFz \propto \exp\left(-\frac{\chi_{\rm min}^2(z)}{2}\right)\ \
 \hbox{with}\ \ \chi^2_{\rm min}(z) = \sum_{i}\left[ {F_{{\rm
 obs},i}-s\cdot F_{{\rm tem},i}(z) \over \sigma_i} \right]^2\ ,
\label{eq:pbz}
\ee

where $\chi_{\rm min}^2(z)$ is the best fit value obtained at redshift
$z$; $F_{{\rm obs},i}$ is the observed flux; $F_{{\rm tem},i}(z)$ is
the template flux at redshift $z$ in $i$-th band, $\sigma_i$ is the
photometric error in $i$-th band and $s$ is the scaling factor applied
to the template fluxes as described in A99 (Equation 2). The $PDFz$ is
then normalised to unity over the full range used to derive the
redshift (here $0\le z
\le 6$).

\begin{figure*}
\centering
\hbox{ 
\subfigure{\psfig{figure=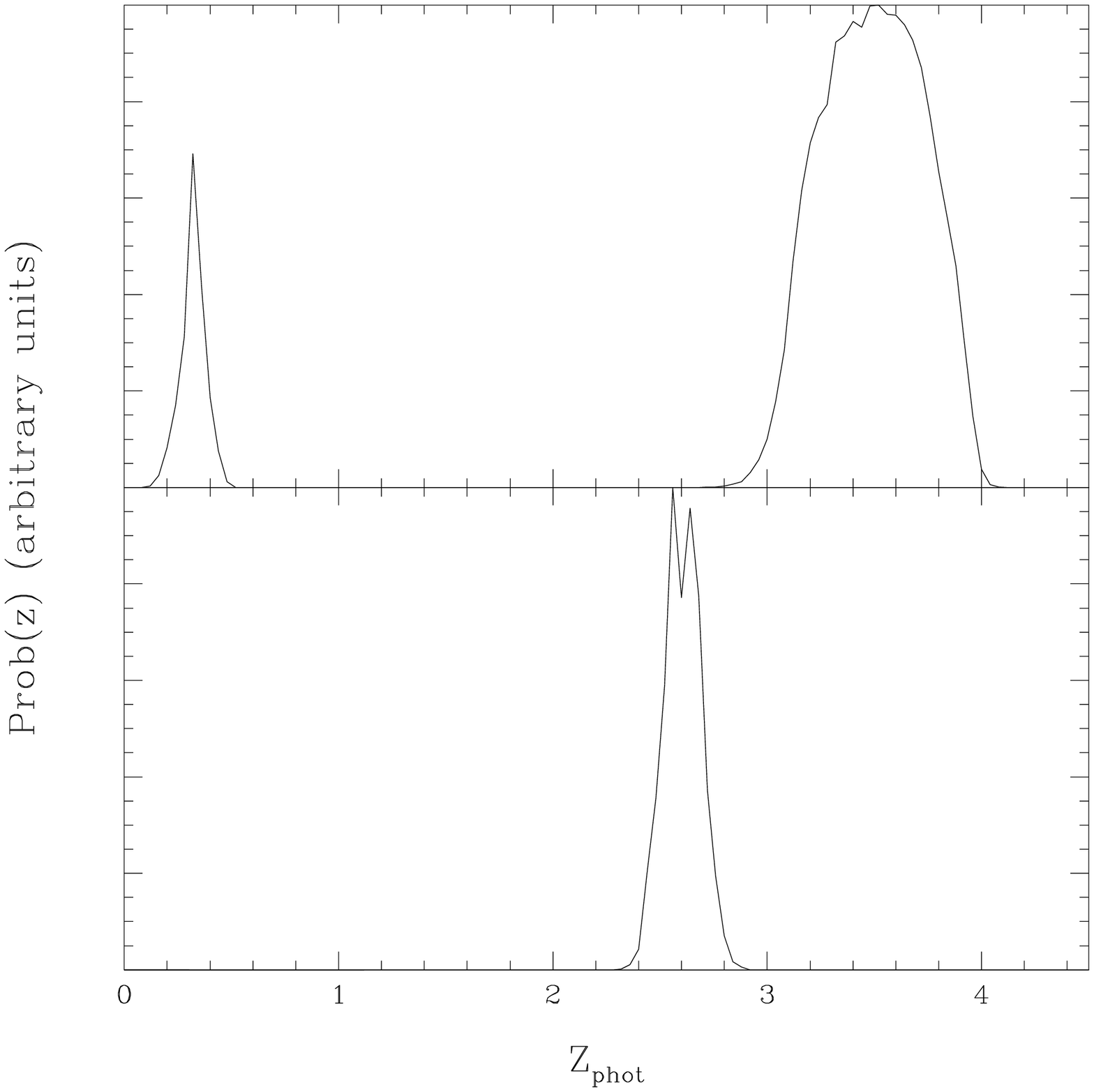,angle=0,width=6.cm}}
\subfigure{\psfig{figure=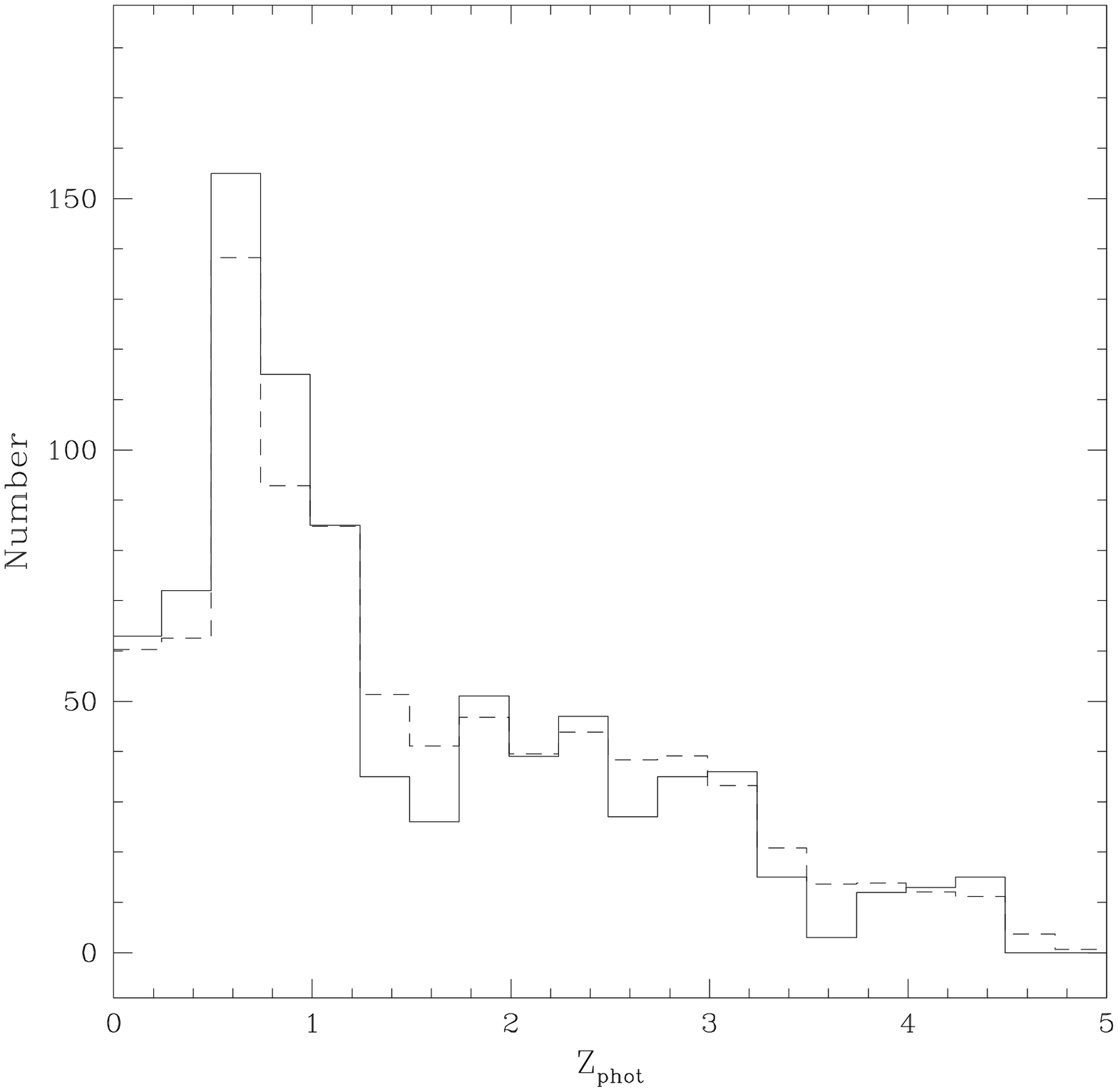,angle=0,width=6.cm}}
	}
\caption[]{ Left panel: Two examples of redshift probability function
 for one object with a secondary redshift peak (upper panel) and for
 one object without secondary peak (lower panel). The area below the
 curves is normalised to unity. Right panel: Redshift distributions for
 galaxies with $I_{AB}\le 27.5$ using the best redshift value for each
 object (solid line histogram) and by summing up the normalized $PDFz$
 (dashed line histogram).  }
\label{fnz}
\end{figure*}

This $PDFz$ makes it possible to follow the redshift probability for
each object (see also Bolzonella, Miralles \& Pell\`o 2000) and has
some similarity with the Bayesian photometric redshift estimation
(Ben\'{\i}tez 2000). To illustrate the behaviour of the $PDFz$, in
Figure~\ref{fnz}, (left panel) we show two examples for one object at
$z_{\rm phot}=3.52$ with a secondary peak at $z_{\rm phot}=0.32$
(upper panel) and one at $z_{\rm phot}=2.56$ with no secondary peak
(lower panel). In Figure~\ref{fnz} (right panel) we also compare the
redshift distribution for objects brighter than $I_{AB}=27.5$ obtained
by using the best redshift for the sources (solid line) and by summing
the normalized $PDFz$ of all objects (dashed line).  The spread in the
individual $PDFz$ results in a sort of smoothing of the distribution
obtained with the best redshift estimates.

%--------------------------------------------------------
%--------------------------------------------------------

\subsubsection{Weighted ACF measurement}

\begin{figure*}
\centerline{\psfig{figure=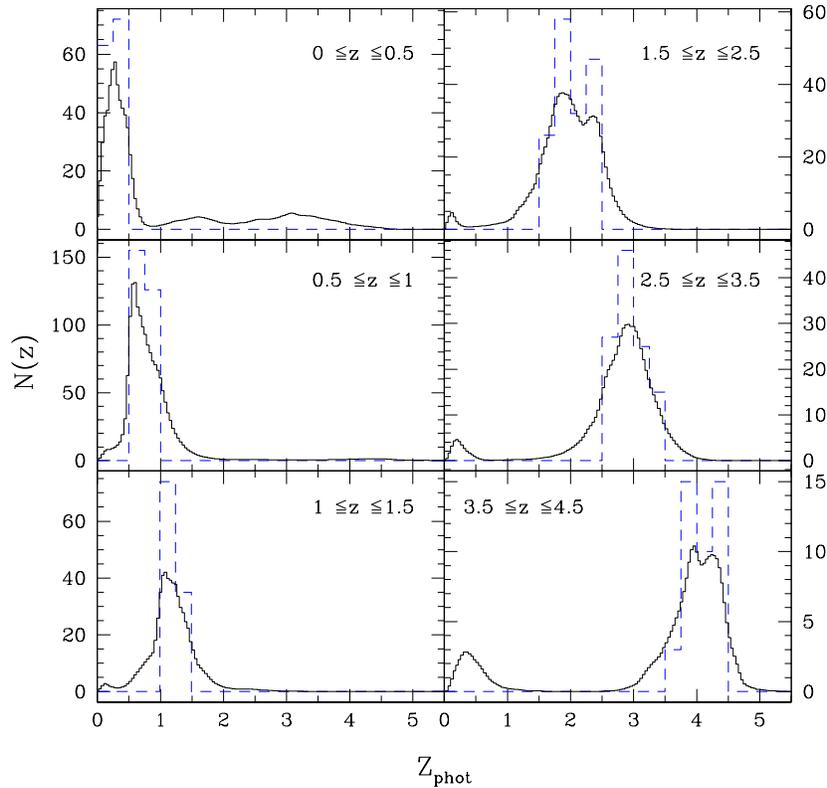,angle=0,width=11.cm}}
\caption[]{Histograms of photometric redshifts in
  different redshift bins (specified in each panel) defined by the
  best redshift (dashed lines) as compared with the histograms
  obtained for the same objects taking into account the redshift
  probability function (solid lines). }
\label{fpdz} 
\end{figure*}

In the previous section, we have computed the ACF assuming the best
redshift value for each object regardless of its confidence level.  In
this section we take the redshift uncertainty into account directly in
the ACF measurement by using the $PDFz$ of each object.  For all the
galaxies within a given redshift interval, we use the $PDFz$ to weight
the number of pairs according to the probability of the objects being 
in the redshift bin.  In Figure~\ref{fpdz} we compare in the different
redshift intervals the distribution obtained with the best redshift
approach and that resulting from the summed $PDFz$ approach. The
results are summarized in Table~\ref{tzcont}. We find that: \\

\noindent $1)$ The summed $PDFz$ are in general similar to the original
 distributions and have tails in the neighbouring intervals.  The
 enlargement of the distribution corresponds to a change between 5 and
 30 per cent.  This effect is mainly due to the redshift uncertainties
 of objects at the boundaries of the bins under consideration. \\

\noindent $2)$ At redshifts between 0.5 and to 3.5, 
a very small fraction of objects shows catastrophic secondary
redshifts ($\le 5$ per cent). For the extreme bins the behaviour is
different. The $3.5\le z\le 4.5$ bin shows a pronounced secondary peak
at low $z$ corresponding to $13.5$ per cent of the total. The $0 \le z
\le 0.5$ redshift bin shows a long tail between $1 \le z \le 4$
corresponding to a fraction of 31.5 per cent.

We find that the fraction of lost objects for different redshift
ranges is in good agreement with the Monte Carlo simulations carried
out in A99, where a gaussian random noise has been added to the
original photometric errors (see Figure 4 of A99). This shows that the
two approachs provide similar results to quantify the photometric
redshift confidence levels.

%--------------------------------------------------------
%--------------------------------------------------------

%--------------------------------------------------------
\begin{table}
\caption[]{Distribution of the $PDFz$ for different redshift intervals.
Column 1: redshift bin.  Column 2: fraction of objects within the
redshift range. Column 3: fraction distributed in the adjacent
bins. Column 4: fraction in non-adjacent bins.  }
\begin{tabular}{cccc}
\hline
  $z$ range  & Fraction in  & Fraction in        &  Fraction in     \\
             & the bin (\%) & adj. bins  (\%) &  non adj. bins  (\%) \\
\hline
0.0  -  0.5     & 63   &  5.5 & 31.5 \\
0.5  -  1.0     & 73.5 & 21.5 &  5   \\
1.0  -  1.5     & 66   & 30   &  4   \\
1.5  -  2.5     & 79   & 17   &  4   \\
2.5  -  3.5     & 78.5 & 16.5 &  5   \\
3.5  -  4.5     & 74.5 & 12   & 13.5 \\
\hline
\end{tabular}
\label{tzcont}
\end{table}
%--------------------------------------------------------
 
Since the ACF measurements in the HDFs is based on a small sample, we
want to optimize the reliability of the signal in each redshift
bin. The strategy adopted in the following analysis is to include in
each redshift bin only the objects for which the best redshift belongs
to the bin. We call $N_{\rm data}$ their number. This allows also a
direct comparison between the classical ACF and this method.

To measure the weighted ACF, the number of pairs in the redshift range
$z_{\rm min} \le z \le z_{\rm max}$, entering in equation~\ref{ew}, is
replaced as follows:
\begin{equation}
DD \ = \ \sum_{i,j}^{N_{\rm data}} Pb^i \cdot Pb^j \hbox{\ ; \ }
DR \ = \ \sum_{i=1,j=1}^{N_{\rm data},N_r} Pb^i\ ,
\end{equation}   
where $Pb^i$ represents the integral of $PDFz$ between $z_{\rm min}$
and $z_{\rm max}$ for the $i$-th object.

The normalisation factors $A_1$ and $A_2$ are the same except that the
total number of objects $n_D$ is replaced by $n_D=\sum_{i=1}^{N_{\rm
data}} Pb^i$.

The results are presented in the next subsection.

%
%%%%%%%%%%%%%%%%%%%%%%%%%%%%%%%%%%%%%%%%%%%%%%%%%%%%%%%%%%%%
%
\subsection{Results}

\begin{figure*}
\centerline{\psfig{figure=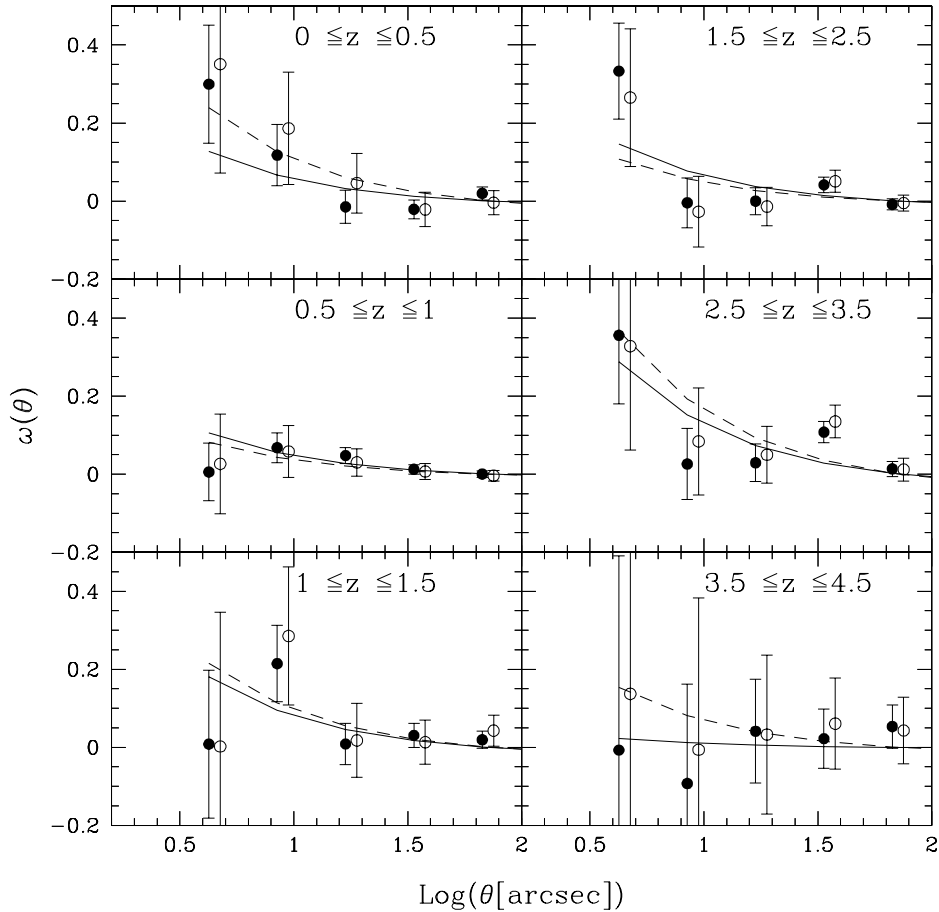,angle=0,width=13.cm}} 
\caption[]{
       The angular correlation functions $\omega(\theta)$ for galaxies
       with $I_{AB}\le 27.5$ measured for different redshift intervals
       (as specified in each panel).  The uncertainties are nearly
       Poisson errors.  The results and the power-law best-fit
       obtained using the classical ACF estimator are shown by filled
       circles and solid lines, while open circles and dashed lines
       refer to the results obtained with the weighted ACF estimator.}
\label{fwdz}
\end{figure*}

In this section we discuss the clustering properties of the HDF-South.
Figure \ref{fwdz} presents the measurements of the ACF in different
redshift bins obtained using both methods discussed in the previous
subsections.  In particular, filled circles refer to the classical ACF
estimates and open circles to the weighted ACF ones.  Note that the
errorbars are slightly larger for this last method. In fact, in this
case the effective number of contributing points $n_D$ is smaller, as
the galaxies have a non-vanishing probability outside their bin.

In order to give a more quantitative estimate of the correlation
strength, we fit the data by adopting a power-law form for the ACF as
$\omega(\theta)=A_{\omega} \theta^{-\delta}$.  If the spatial
correlation function $\xi$ is also assumed to follow a power-law
relation, i.e. $\xi(r)=(r/r_0)^{-\gamma}$, the slope $\gamma$ is
simply related to $\delta$: $\gamma=\delta+1$.  Since the galaxy
samples are small, we prefer to derive the amplitude $A_{\omega}$ by
fixing the value of $\delta$.  As in A99, we adopt $\delta=0.8$ but we
will discuss this assumption later.

Due to the small size of the considered field, we have to take into
account the integral constraint $IC$ (Peebles 1974) in our fitting
procedure as:
\be 
\omega_{\rm est} \simeq\omega_{\rm true} - IC \ .
\ee
The quantity $IC$ is defined as the integral of the ACF over the
survey, i.e.
\be 
IC = \overline{\omega}_{\theta_{\rm max}} =
\frac{1}{\Omega^2} \int \int \omega(\theta) d\Omega_1 d\Omega_2 =
A_{\omega} \times B\ ,
\label{eic}
\ee
where $\theta_{\rm max}$ is the maximum scale of the survey.  The
integral $B$ has been computed by a Monte-Carlo method using the same
geometry as the HDF-South and masking the excluded regions.  Adopting
the value $\delta=0.8$, we derive $B = 0.033$ (for $\theta$ measured
in arcsec).

The fitting power-law relations are all shown in Figure
\ref{fwdz}, both for the classical ACF (solid lines) and weighted ACF
(dashed lines), while the values of the amplitude of $\omega(\theta)$
at 10 arcsec are reported in Table~\ref{tas}.  In general we find a
good agreement between the results of the two different techniques.
We find some differences only in the two extreme redshift bins ($
\langle z \rangle =0.25$ and $\langle z \rangle =4$), where the
objects typically display significant tails in the $PDFz$.  Here the
weighted ACF seems to allow a better extraction of the signal, giving
larger values for the correlation function. However, due to the large
errorbars, the two methods are still consistent at the $1\sigma$
level. Finally we note that in the redshift bin between $3.5 \le z \le
4.5$ the results are consistent with the assumption of vanishing
clustering.

%
%--------------------------------------------------------
%--------------------------------------------------------
%

In order to discuss the effects of the assumed slope on the clustering
normalisation, in Table~\ref{tas} we also show the amplitudes obtained
using $\delta=0.6$ and $\delta=0.9$.  The integral constraints $IC$
have been recomputed according to the slope: we find $B=0.074$ and
$B=0.022$ for $\delta=0.6$ and $\delta=0.9$, respectively.  The
results show that the impact of the changes in the assumed slope
affects the values of $A_\omega$ at 10 arcsec by less than $ 1
\sigma$.

%
%-----------------------------------------------------------
\begin{table}
\caption[]{ The amplitude of $\omega(\theta)$ at 10 arcsec ($A_\omega$)  
for different redshift bins. Column 1: redshift interval. Column 2:
number of galaxies with $I_{AB}\le 27.5$ and best photometric redshift
belonging to the redshift bin.  Columns 3 and 5: amplitude
$A_{\omega}$ computed assuming a slope $\delta=0.8$ for the classical
and weighted ACF estimator, respectively.  Column 4: amplitude
$A_\omega$ computed using the classical ACF but assuming a different
slope ($\delta=0.6$ and $\delta=0.9$). }
\begin{tabular}{|c|c|cc|c|} \hline
 &  & \multicolumn{2}{|c|}{Classical ACF} &  
\multicolumn{1}{|c|}{Weighted ACF} \\ \hline   
 $z$ range  & Number  & \multicolumn{2}{|c|}{$A_{\omega}$(10arcsec)} 
 & $A_{\omega}$(10arcsec) \\
 & $I_{AB}\le 27.5$ &  $\delta=0.8$ & $\delta=0.6, 0.9$   & $\delta=0.8$  
\\  \hline
0.0 - 0.5     &  135 & 0.07$\pm$0.05 & 0.07,0.07 & 0.13$\pm$0.08 \\
0.5 - 1.0     &  281 & 0.06$\pm$0.02 & 0.07,0.06 & 0.05$\pm$0.04 \\
1.0 - 1.5     &  109 & 0.10$\pm$0.06 & 0.12,0.09 & 0.12$\pm$0.09 \\
1.5 - 2.5     &  163 & 0.08$\pm$0.04 & 0.09,0.08 & 0.06$\pm$0.05 \\
2.5 - 3.5     &  113 & 0.16$\pm$0.05 & 0.19,0.15 & 0.21$\pm$0.08 \\
3.5 - 4.5     &   43 & 0.01$\pm$0.15 & 0.03,0.00 & 0.09$\pm$0.23 \\
\hline
\end{tabular}
\label{tas}
\end{table}
%--------------------------------------------------------
%

\begin{figure*}
\centering
\hbox{ 
\subfigure{\psfig{figure=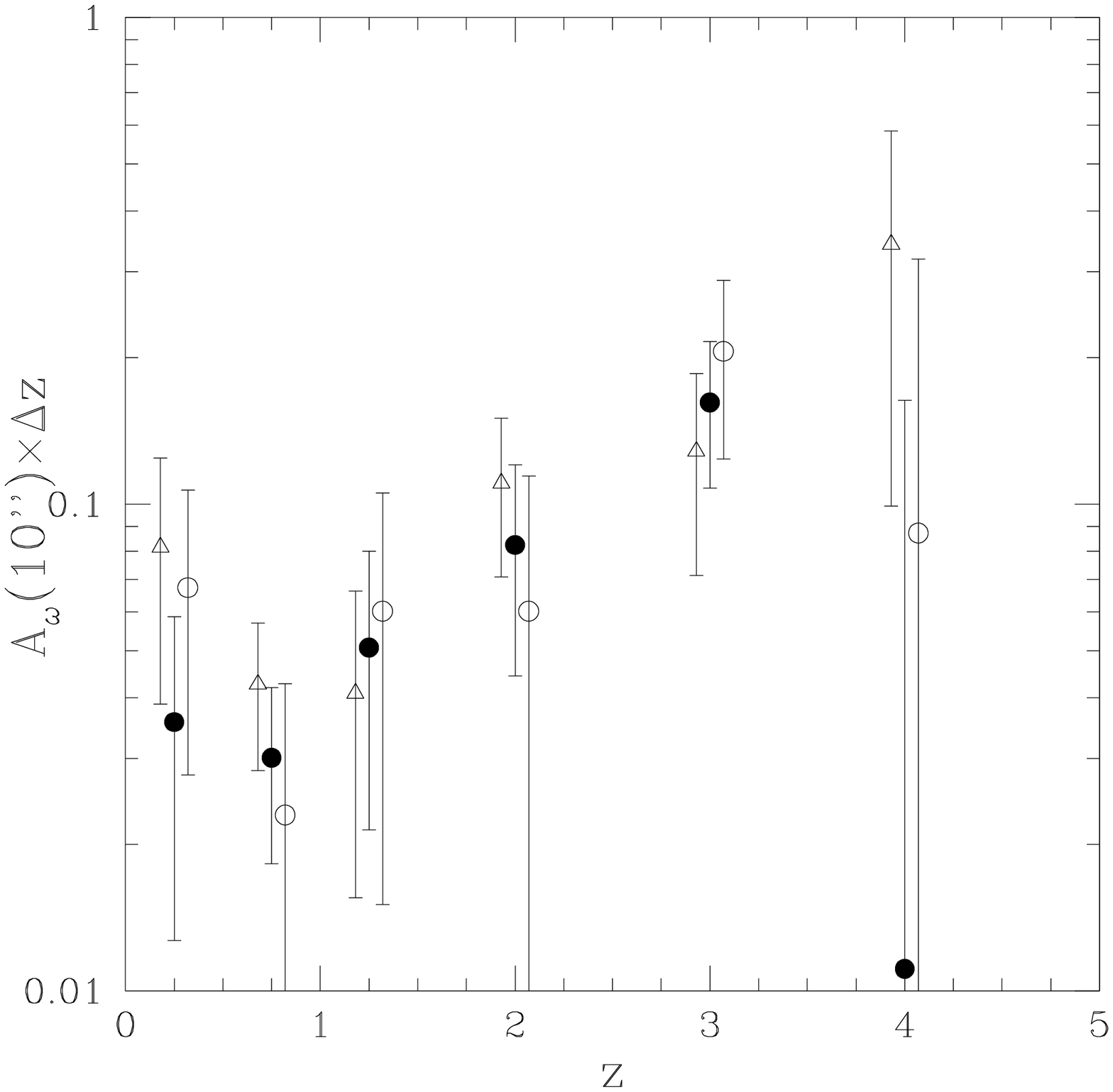,angle=0,width=7.5cm}}
\subfigure{\psfig{figure=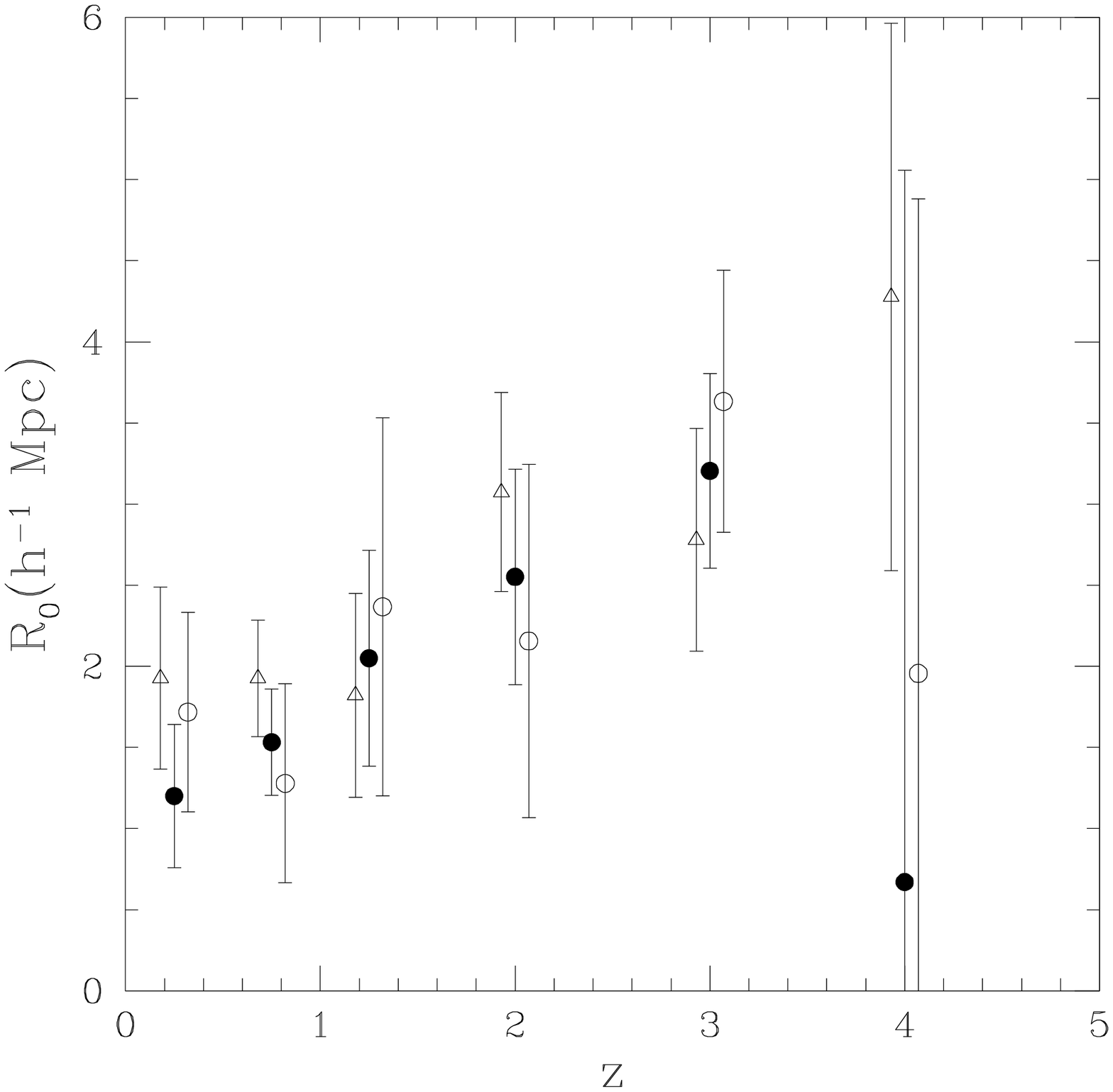,angle=0,width=7.5cm}}
     }
\caption[]{Comparison of the clustering properties of galaxies 
in HDF-North and South. Left panel: the redshift evolution of the ACF
amplitude $A_\omega$ at 10 arcsec (multiplied by the bin size $\Delta
z$).  Open triangles refer to the values of the HDF-North obtained by
A99, while filled and open circles refer to the results obtained in
this work adopting the classical and the weighted estimators,
respectively.  The different measurements have been shifted by
$z=\langle z \rangle +/-0.05$ for clarity.  Right panel: the redshift
evolution of the comoving correlation length $r_0(z)$ (in $h^{-1}
Mpc$) as computed by assuming a flat universe with $\Omega_{\rm
0m}=0.3$.  The meaning of different symbols is the same as in the left
panel. }
\label{fAwdz}
\end{figure*}

It is important now to compare the clustering properties of HDF-South
to the corresponding results for HDF-North that we obtained in our
previous analysis (A99). The comparison is presented in
Figure~\ref{fAwdz}.  The left panel shows the behaviour of
$A_{\omega}$ computed at 10 arcsec (and multiplied by the bin size
$\Delta z$, for consistency with A99). In spite of the smallness of
the regions, the amplitudes of the correlation function measured in
the two Hubble deep fields are in good agreement, showing a small
field-to-field variation. The results confirm the behaviour of the
clustering with the redshift we found in A99. Namely, the clustering
amplitude declines from $z=0$ to $z\sim 1$ and increases at higher
redshifts to become, at $z \ge 2$, comparable to or higher than that
observed at $z\simeq 0.25$.  At $z\simeq 4$ the clustering signal
measured in HDF-South is very noisy and we cannot confirm the high
value of $A_\omega$ found in the northern field.  An alternative
measure of the correlation strength is the comoving correlation length
$r_0$. Its redshift evolution, computed, as in Magliocchetti
\& Maddox (1999), assuming a flat universe with present matter density 
parameter $\Omega_{\rm 0m}=0.3$, is shown in the right panel of
Figure~\ref{fAwdz}.  Again, we find a slightly declining or almost
constant behaviour up to $z\simeq 1$ and an increasing trend from
$z\simeq 2 $ to $z \simeq 3$.

%
%--------------------------------------------------------
%--------------------------------------------------------
%

\begin{figure*}
\centerline{\psfig{figure=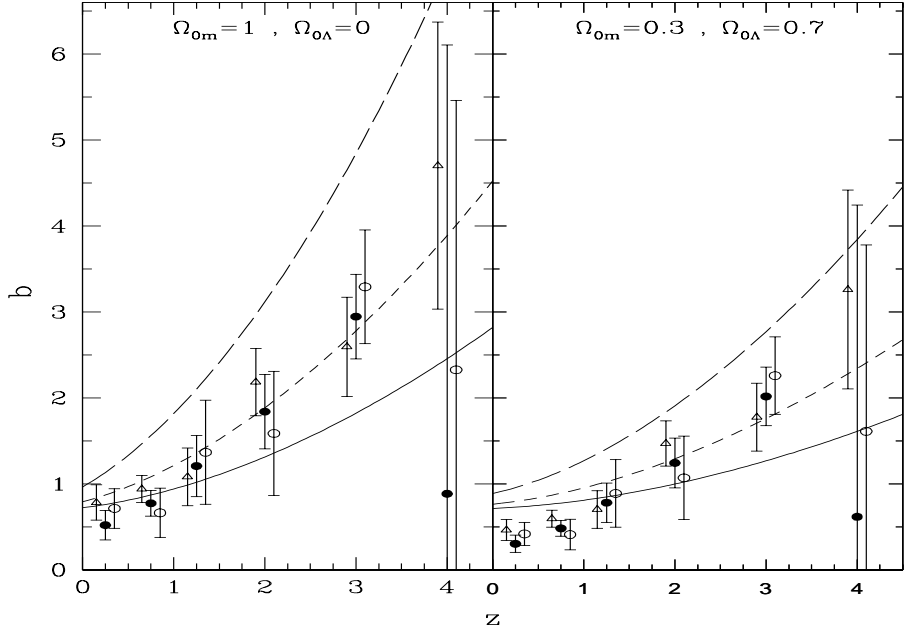,angle=0,width=13cm,height=9cm}}
       \caption[]{ The measured bias $b$ as a function of redshift for
       an Einstein-de Sitter $SCDM$ model and $\Lambda CDM$ model
       (left and right panels).  The open triangles refer to the
       values obtained for the HDF-North in A99; filled and open
       circles refer to the values obtained in this work for the
       HDF-South using the classical ACF and the weighted ACF
       estimators, respectively.  The different lines represent the
       theoretical effective bias computed for the same cosmological
       models assuming different values of minimum mass $M_{\rm min}$.
       We show results for $M_{\rm min}=10^{10}$ (solid lines),
       $10^{11}$ (short-dashed lines) and $10^{12}$ (long-dashed
       lines) $h^{-1} M_{\odot}$.}
\label{fbias}
\end{figure*}

Since the clustering amplitude of the dark matter decreases
continuously with redshift (the actual behaviour depending on the
cosmological scenario), the observed increase of the galaxy clustering
at high redshift implies that galaxies at $z\approx 3-4$ are biased
tracers of the underlying dark matter.  This effect is illustrated in
Figure~\ref{fbias}, where we show the bias parameter $b$ as a function
of redshift both for the HDF-South and HDF-North.  The values of $b$
are computed by dividing the rms galaxy density fluctuation inside a
sphere of $8 h^{-1} Mpc$ at a given $z$ ($\sigma_8^{\rm gal}$) by the
rms mass density fluctuation ($\sigma_8^{\rm m}$) predicted by linear
theory.  We consider two cosmological models with a cold dark matter
($CDM$) power spectrum normalised to reproduce the local cluster
abundance (Eke, Cole \& Frenk 1996): an Einstein-de Sitter $SCDM$
model ($\sigma_8^{m}(z=0)=0.52$ and $\Gamma=0.45$; left panel) and a
flat $\Lambda CDM$ model with $\Omega_{\rm 0m}=0.3$ and
$\Omega_{0\Lambda}=0.7$ ($\sigma_8^{m}(z=0)=0.93$ and $\Gamma=0.21$;
right panel).  The observed bias parameters for the HDF-South are in
good agreement with our previous results for the northern field. In
particular we observe some anti-bias ($b(z<1)\sim 0.5$) at low
redshift, while we confirm that the high-redshift galaxies are
strongly biased with respect to the dark matter: $b(z\sim 3)\sim 3 ,
2$ for the $SCDM$ and $\Lambda CDM$ models, respectively.  This
supports a model of biased galaxy formation where $b$ is evolving with
redshift.  For comparison, in the same plot we also show the
theoretical expectations for the effective bias (see Matarrese et
al. 1997 and Moscardini et al. 1998 for a definition) computed for the
same cosmological models using different minimum mass for the dark
matter haloes ($M_{\rm min}=10^{10}, 10^{11},10^{12} h^{-1}
M_{\odot}$). For the Einstein-de Sitter model, we can reproduce the
observations with a minimum mass $M_{\rm min} \simeq 10^{10} h^{-1}
M_{\odot}$ at $z\le 1$ and $M_{\rm min} \simeq 10^{11} h^{-1}
M_{\odot}$ between $1\le z \le 3$.  For the $\Lambda CDM$ model
$M_{\rm min} < 10^{10} h^{-1} M_{\odot}$ is required at $z\le 1$ ,
$10^{10}\le M_{\rm min} \le 10^{11} h^{-1} M_{\odot}$ for $1\le z \le
2$ and $M_{\rm min} \ge 10^{11} h^{-1} M_{\odot}$ at $<z>=3$ are
required.

At redshift $z\simeq 3$, alternative estimates of the galaxy
clustering come from the analysis of the Lyman Break Galaxy (LBG)
samples.  We find that the bias measured for the HDF-population is
smaller than the one observed for the bright LBGs (Steidel et
al. 1996).  Assuming for example an Einstein-de Sitter model,
Adelberger et al. (1998) found for the spectroscopic sample of LBGs a
bias parameter $b(z=3)\approx 6$ while Giavalisco et al.  (1998) found
$b(z=3)\approx 4.5$ from the photometric sample (see also Giavalisco
\& Dickinson 2001).  Averaging our results for HDF-South and
HDF-North, we find $b(z=3)\approx 2.8$.  These differences can be
explained by the different surface galaxy densities (larger in the HDF
fields, which have approximately 30 objects per square arcmin).  In
fact in the hierarchical galaxy formation scenario, more massive and
rare objects form in rarer and higher peaks of the underlying matter
density field; as a consequence they are expected to have a higher
value of the bias parameter.
 
\section{The error budget}
\label{s:errors}

Up to now, we have assumed in our measurements that the errors are
nearly Poissonian.  We have neglected other possible contributions to
the errors, due to the finite area of the survey or to the clustered
nature of the galaxy distribution. In this section we deal with these
effects basing our analysis on analytical expressions of the cosmic
errors calculated by Bernstein (1994).  We estimate the relative
magnitude of various contributions to the errors and show that in fact
our nearly Poissonian errorbars are consistent.  We also examine the
possible improvements brought by a survey made with the Advanced
Camera on HST.

\subsection{Analytic expression for cosmic uncertainties}   

Originally, LS93 have derived the variance of their estimator by
assuming the weak correlation limit but neglecting the contribution of
the higher-order correlation functions. The computation has been
generalized by Bernstein (1994, hereafter B94; see also Hamilton 1993;
Szapudi 2000) for any clustering regime taking into account
higher-order correlation functions but neglecting edge effects.  The
Bernstein's equation is obtained in the case of a (degenerate)
hierarchical model and has been rewritten as follows:

%%%%%%%%%%%%%%%%%%%%%%%%%%%%%%%%%%%%%%%%%%%%%%%%%%%%%%%%%%%%%%%%%%%%%
\begin{eqnarray}
\left(\frac{\Delta\omega(\theta)}{\omega(\theta)}\right)^2
 & \simeq & 4\ (1-2q_3 + q_4)\
 \frac{\overline{\omega}_{\theta_{\rm max}}}{\omega(\theta)^2} \
 (\omega(\theta) -\overline{\omega}_{\theta_{\rm max}})^2 \ + \
\left(\frac{\overline{\omega}_{\theta_{\rm max}}}{\omega(\theta)}\right)^2 
 \hspace{3.cm} (E_1) \nonumber \\
  &+& \frac{4}{N_g}\ \left[ \frac{\omega_r(\theta)(1+2q_3 
  \omega(\theta))}{\omega(\theta)^2} +q_3 -1 \ 
 + \ \frac{\overline{\omega}_{\theta_{\rm max}}}{\omega(\theta)^2}
\left[ 2(1-2q_3)\omega(\theta) \ - \ 
 2q_3\omega_r(\theta)+\overline{\omega}_{\theta_{\rm max}}(3q_3-1) -1
\right]\right]   
 \hspace{0.2cm}  (E_2) \nonumber  \\
  &+& \frac{2}{N_g^2}\ \left[ (G_p(\theta)^{-1}-1)\frac{1+\omega(\theta)}{\omega(\theta)^2} 
 -1 -\frac{1-2\overline{\omega}_{\theta_{\rm max}}}{\omega(\theta)}\ + \ 
\frac{\overline{\omega}_{\theta_{\rm max}}}{\omega(\theta)^2}\left(\frac{2}{G_p(\theta)} - 1 - \overline{\omega}_{\theta_{\rm max}} \right)
 \right].
\hspace{0.2cm}  (E_3) \nonumber  \\
\label{ece}
\end{eqnarray}
%%%%%%%%%%%%%%%%%%%%%%%%%%%%%%%%%%%%%%%%%%%%%%%%%%%%%%%%%%%%%%%%%%%%%

In this equation, valid in the regime $\theta/\theta_{\rm max}$ ,
$\overline{\omega}_{\theta_{\rm max}}$, $1/N_g \ll 1$, $N_g$ is the
number of galaxies in our sample.  The function $\omega_r(\theta)$ is
the average of the two-point correlation over a shell corresponding to
angles in the bin $[\theta, \theta+\delta\theta]$.  In a first
approximation, $\omega_r(\theta) \simeq \omega(\theta)$ (see B94).
The term $G_p(\theta)$ is the probability of finding two randomly
placed galaxies with separation in the range $[\theta,
\theta+\delta\theta]$:
\be
G_p(\theta) = < RR(\theta)> / [N_g (N_g-1)/2]. 
\ee

The parameters $q_3$, $q_4$ are related to the hierarchical amplitudes
of the cumulants of the dark matter distribution $S_3$, $S_4$ ($S_N
\equiv {\bar \omega}_N > / [{\bar\omega}]^{N-1}$, where $\omega_N$ are
the $N$-point angular correlation functions, and ${\bar \omega}_N$
corresponds to their integral average over a disk of radius $\theta$)
by $q_3\simeq S_3/3$ and $q_4\simeq S_4/16$. 

Equation~(\ref{ece}) is composed of three terms, which we call $E_1$,
$E_2$ and $E_3$.

The first contribution to the errors, $E_1$, hereafter referred as the
finite volume error\footnote{also often called ``{\it cosmic
variance}''} (e.g. Szapudi \& Colombi 1996), does not depend on the
number of galaxies in the catalogue. It comes from the finiteness of
the area covered by the survey. In a first approximation, this is
proportional to the average of the two-point correlation function over
the survey area, $\overline{\omega}_{\theta_{\rm max}}$ [see
equation~(\ref{eic})].

The second ($E_2$) and third ($E_3$) terms reflect the discrete nature
of the catalogue. They account for random fluctuations of the galaxy
distribution as a local Poisson realization of a continuous underlying
field (e.g. Szapudi \& Colombi 1996). The term $E_2$, proportional to
$1/N_g$, appears only in correlated sets of points (see B94): it
cancels in the Poisson limit, $\omega \longrightarrow 0$. The pure
Poisson error is in fact contained in the next order term, $E_3$,
proportional to $1/N_g^2$.  Hereafter, $E_2$ and $E_3$ will be
referred to as the discreteness errors.  Note that discreteness and
finite volume effects can be disentangled only approximately: there
are terms proportional to $\overline{\omega}_{\theta_{\rm max}}$ in
$E_2$ and $E_3$. They correspond to hybrid, ``finite-discreteness''
effects. However these latter give only very little contribution to
$E_2$ and $E_3$ and can in fact be neglected in most realistic
situations.  Finally, the error estimate of B94 neglects edge effects
which become significant at the largest angular scales. The advantage
of the LS estimator is to reduce these latter as much as possible, and
therefore equation~(\ref{ece}) is expected to give a good estimate of
the cosmic errors even in this regime, although it might slightly
underestimate them.

\subsection{Assumptions used to compute the cosmic errors}
\label{s:ototot}
 From equation~(\ref{ece}), one can see that the calculation of the
cosmic error for $\omega(\theta)$ requires prior knowledge of
statistics up to order four, in particular $\omega(\theta)$ itself,
$\overline{\omega}_{\theta_{\rm max}}$, $q_3$ and $q_4$. To estimate
them, we proceed as follows.
\begin{itemize}
\item The value of $\omega(\theta)$ taken in equation~(\ref{ece}) is
estimated from the best fits, $\omega_{\rm fit}$, obtained in
Figure~\ref{fwdz};
\item The calculation of $\overline{\omega}_{\theta_{\rm max}}$ is done as
explained at the end of \S~\ref{s:cacf}. Note that computing the
integral constraint in such a way, by assuming a power-law behavior
for the two-point correlation function in all the regimes, might in
turn lead to overestimating $\overline{\omega}_{\theta_{\rm
max}}$. Indeed $\omega(\theta)$ is expected to present a cut-off at
large scales, at least if low-$z$ results (such as measurements of
$\omega(\theta)$ in the APM; e.g. Maddox et al. 1990) can be
extrapolated to higher redshifts.
\item The choice of $q_3$ and $q_4$ is more delicate: these parameters
cannot be inferred from self-consistent measurements in the catalogues
analysed in this paper and in A99. Indeed, higher-order statistics are
more sensitive to cosmic errors than the two-point correlation
function, with an error which increases with the order
considered. Therefore it would be impossible to extract reliable
values of $q_3$ and $q_4$ from these catalogues mainly contaminated by
shot-noise, even with strong prior assumptions such as assuming a power-law
behaviour for higher-order correlation functions similarly as we did
for $\omega(\theta)$.  Instead, we use measurements of $S_3=3\, q_3$
and $S_4=16\, q_4$ obtained in the local universe ($z=0$) by
Gazta\~naga (1994) with the APM catalogue (Maddox et al. 1990) at
$\theta\sim 0.1^{\circ}$: $S_{3}(z=0)\simeq 4$ and $S_{4}(z=0)\simeq
50$. At the level of approximation used in this paper, we can neglect
a possible dependence of $S_3$ and $S_4$ on the angular
scale. However, evolution with redshift of these quantities might be
important, particularly if the bias between the galaxy and the dark
matter distributions increases significantly with redshift, as
suggested by the measurements in this paper.  Both theoretical
calculations based on perturbation theory (e.g. Juszkiewicz, Bouchet
\& Colombi 1993; Bernardeau 1994) and measurements in N-body
simulations (e.g. Colombi, Bouchet \& Hernquist 1996; Szapudi et
al. 1999) show that the parameters $S_3$ and $S_4$ measured in the
dark matter distribution do not evolve significantly with time, at
least at the level of approximation of this paper. However, the bias
can strongly affect higher-order statistics: in general, increasing
the bias factor $b$ reduces the values of $S_3$ and $S_4$ compared to
what is obtained in the dark matter distribution.  Here, following
Colombi et al. (2000), we adopt two simple, extreme models. The first
one consists in assuming that the effect of biasing is negligible:
$S_N(z)=S_N(z=0)$, that we refer to as the no bias model.
 The second one is motivated by observational results
(Szapudi et al. 2001) and to some extent by theoretical calculations
(Bernardeau \& Schaeffer 1992; 1999):
$S_N(z)=S_N(z=0)/[b(z)]^{2(N-2)}$. For the values of $b(z)$, we take the ACF
measurements in the HDF-South obtained with the classical approach
(unless otherwise specified) as shown for each cosmology in
Figure~\ref{fbias} (filled circles) -- i.e. we assume that the APM
galaxies are unbiased with respect to the dark matter distribution and
that the HDF galaxies are biased with respect to the APM ones with
bias equal to $b(z)$.
 These models are referred to as the $SCDM$ and $\Lambda CDM$ bias models.

\end{itemize}

\subsection{The cosmic errors in the HDF fields }

\begin{figure*}
\centerline{\psfig{figure=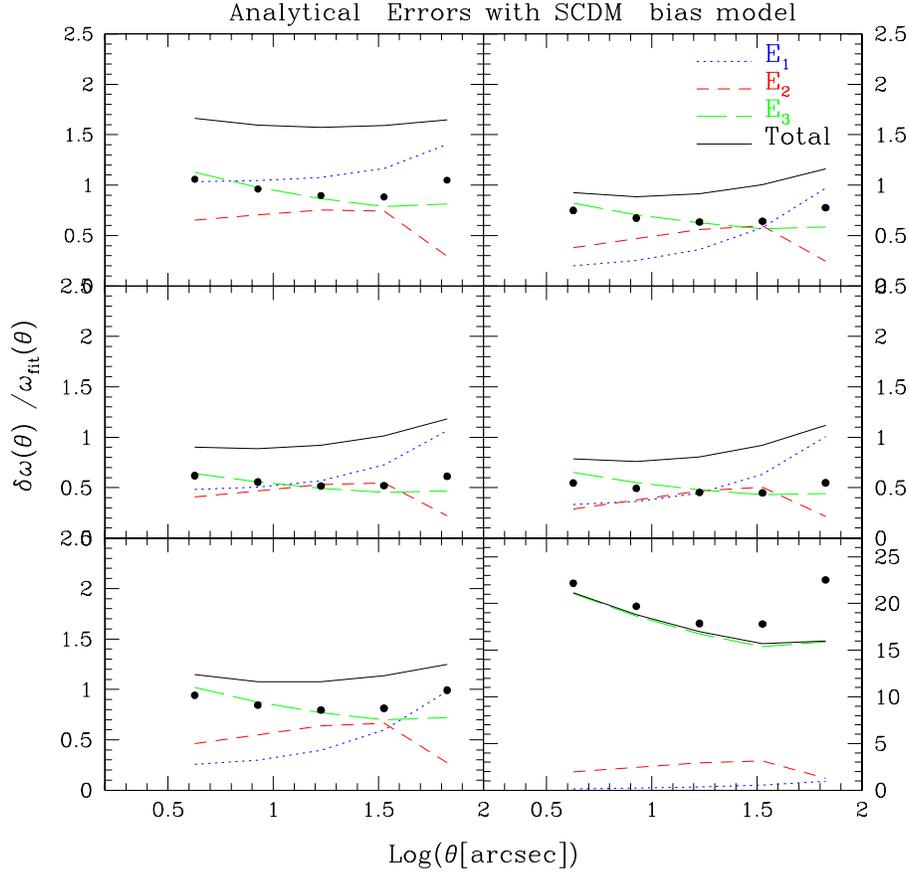,angle=0,width=12cm}}
\caption[]{Comparison between the nearly Poissonian errorbars
[equation~(\ref{ewterr}), filled circles] used in the computation of
$\omega(\theta)$ (Classical ACF) at different angular separations with
the analytical errors of equation~(\ref{ece}): finite volume error
$E_1^{1/2}$ (dotted line); discreteness errors $E_2^{1/2}$
(short-dashed line), $E_3^{1/2}$ (long-dashed line) and total cosmic
error $E^{1/2}\equiv (E_1+E_2+E_3)^{1/2}$ (solid line).  The
analytical errors are computed using the $SCDM$ bias model (see
\S~\ref{s:ototot} for details). The different panels correspond to 
 the different redshift ranges as in Figure~\ref{fwdz}.}
\label{fwt_err}
\end{figure*}

In Figure~\ref{fwt_err}, we compare the magnitudes of the finite
volume error, $E_1^{1/2}$, the discreteness errors, $E_2^{1/2}$ and
$E_3^{1/2}$ [e.g. equation~(\ref{ece})], and the total error $\delta
\omega/\omega_{\rm fit}\equiv E^{1/2}=(E_1+E_2+E_3)^{1/2}$, at
different angular separations with the errors used for the classical
ACF measurement derived from equation~(\ref{ewterr}).  The different
panels show the relative errors for the different redshift ranges as
in Figure~\ref{fwdz}.  We show here the errors obtained from the
analytical expressions using the $SCDM$ bias model ({\it e.g.} $b(z)$
obtained from the left panel of Figure~\ref{fbias}).

As expected, the estimates of the errors used for the classical ACF
[equation~(\ref{ewterr})] match quite well with the $E_3$ term of
equation~\ref{ece} (long-dashed lines).  Because the sample is quite
sparse, we have $E_3 \ga E_2$, but $E_2$ (short-dashed lines) is not
negligible, except at the largest angular separation. The finite
volume error ($E_1$ term, dotted lines) plays an important role as
well, especially at low $z$, where the effective size of the survey is
small, and at large angular scales. Note that the results obtained at
the largest scales have to be interpreted with caution since
equation~(\ref{ece}), which assumes $\theta$ small compared to the
survey size, might be slightly outside its domain of validity.  The
total theoretical cosmic error (solid line) depends weakly on the
scale and assumes its largest values at low and high redshifts: in the
first case because of the finite volume effects, in the second case
because of the Poisson noise.
 
\begin{figure*}
\centerline{\psfig{figure=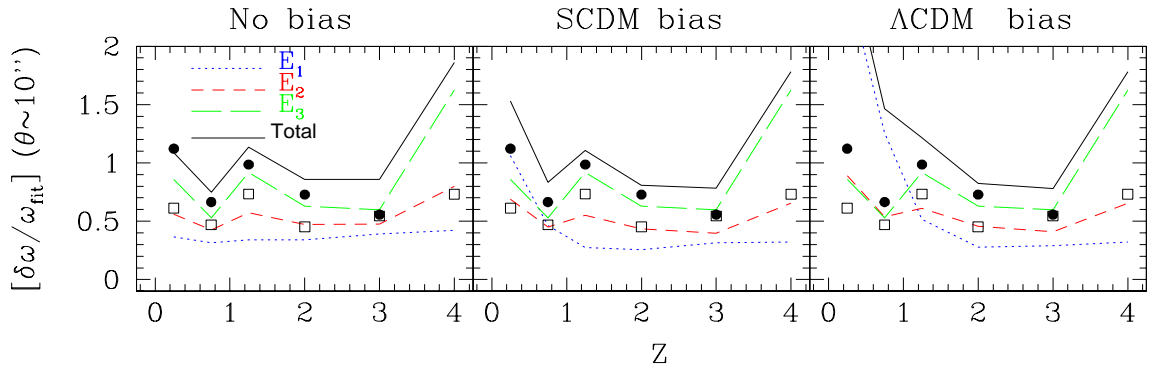,angle=270,width=16cm}}
\caption[]{Comparison of the nearly Poissonian errorbars
(equation~(\ref{ewterr}) at 10 arcsec, estimated for the HDF-South
(filled circles) and the HDF-North (open squares) with the cosmic
errors from equation~(\ref{ece}) (total: solid lines; $E_1$: dotted
lines; $E_2$: short-dashed lines; $E_3$: long-dashed lines).  The
left, middle and right panels refer to the no bias,
the $SCDM$ bias and the $\Lambda CDM$ bias models, respectively
(see text). There is one filled circle missing on each panel at $z=4$,
which corresponds to a very large Poisson errorbar $\delta
\omega/\omega_{\rm fit} \sim 23$ (see the right bottom panel of
Figure~\ref{fwt_err}).}
\label{fawz_err}
\end{figure*}

Figure~\ref{fawz_err} is similar to Figure~\ref{fwt_err}, but shows
the dependence on redshift of the errors at a fixed angular scale,
$\theta=10''$ (this choice being arbitrary). Here, we consider various
bias models: no bias (left panel), the $SCDM$ bias model (middle
panel) and the $\Lambda CDM$ bias model (right panel).  Since the
theoretical expression (equation~\ref{ece}) is now compared to the
equation~\ref{ewterr} for both analyses of the HDF-South (filled
circles) and the HDF-North (open squares, from A99), for any
survey-dependent quantity in equation (\ref{ece}) (namely $N_g$,
$\omega_{\rm fit}(\theta)$ or function $b(z)$), we take the result
obtained from the average between the two fields.

Again, the overall agreement between the $E_3$ term and the errors
estimated from equation~\ref{ewterr} is pretty good, as expected. In
most cases, the term $E_3$ dominates the total error at this angular
scale, except at low $z$ for the bias models, where the finite volume
error dominates. Indeed, as a result of our rather extreme modeling of
the effect of the bias on higher-order statistics, $S_3 \propto
b^{-2}$ and $S_4 \propto b^{-4}$ (\S~\ref{s:ototot}), the effects of
changing $b(z)$ can be important on the finite volume errors,
especially if $b(z)<1$. This is the case at low redshifts in the HDF
population both for $SCDM$, where $b(z=0) \sim 0.7$, and $\Lambda
CDM$, where $b(z=0) \sim 0.4$ (e.g. Figure~\ref{fbias}).

The results presented in Figures~\ref{fwt_err} and \ref{fawz_err}
show that the total theoretical cosmic error given by
equation~(\ref{ece}) can be significantly larger than the estimate
given by equation~(\ref{ewterr}).  This might sound surprising,
because the measured values of $\omega$ are rather small,
$\omega(\theta) \la 0.4$ (e.g Figure~\ref{fwdz}): thus one might argue
that the weak clustering regime approximation (\ref{ewterr}) should be
valid to estimate the errors. In practice, we see that this assumption
is incorrect, particularly at low $z$, at least in the examples
examined here.  However, the amplitude of $E^{1/2}(\theta)$ is at most
$\sim$ twice larger than the error given by equation~(\ref{ewterr})
and shows the same global shape.  Furthermore, as mentioned in
\S~\ref{s:ototot}, ${\bar \omega}_{\theta_{\rm max}}$ is likely to be
overestimated with the method we use, which might artificially
increase the observed difference between equation~(\ref{ewterr}) and
equation~(\ref{ece}).  Finally, one has to be aware of the fact there
is a subtle difference between the calculations of LS93, which lead to
equation~(\ref{ewterr}) and those of Bernstein, which lead to
equation~(\ref{ece}). In the first case, the authors considered a
conditional statistical average, using the supplementary information
that the number of objects in the catalogue $N_g$ is known. In the
second case, the author did not use such information, which naturally
leads to slightly larger errors, since $N_g$ is not conditionally
fixed and can fluctuate.

Given the level of approximation used in this paper, it is thus fair
to conclude that the weak clustering approximation is good enough,
which confirms {\em a posteriori} the validity of the approach used in
A99 and up to \S~\ref{s:errors} in this work, to compute errors.

%
%--------------------------------------------------------
%--------------------------------------------------------
%

%
%--------------------------------------------------------
%--------------------------------------------------------
%

\begin{figure*}
\centerline{\psfig{figure=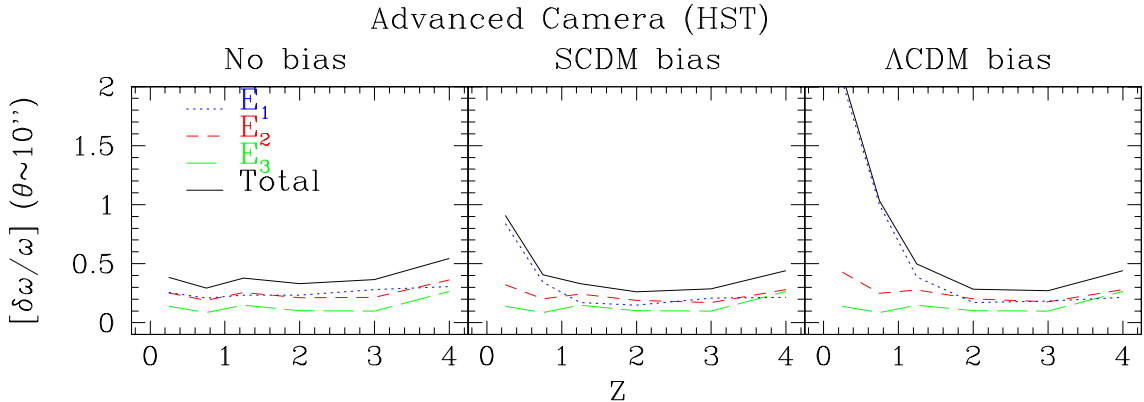,angle=270,width=16cm}}

\caption[]{ As Figure~\ref{fawz_err}, but  in the case 
 of a survey with the Advanced Camera (see text). Only the theoretical
 cosmic errors from equation~(\ref{ece}) are displayed.  }
\label{ce_fov}
\end{figure*}

In order to quantify how the Advanced Camera on the HST (Pirzkal et
al. 2001) can improve the clustering measurements of HDF-like
populations, we have estimated the analytical behaviour of the cosmic
errors with redshift.  The results are shown for $\theta=10''$ in
Figure~\ref{ce_fov}, which can be directly compared with
Figure~\ref{fawz_err}.  To estimate the cosmic errors in
Figure~\ref{ce_fov}, we have rescaled the mean observed number of
galaxies in the HDF-South and North to the respective area of the
Advanced Camera (i.e.~by a factor $5.33$) and recomputed the term
$\overline{\omega}_{\theta_{\rm max}}$ according to the new area,
assuming a square geometry.  Other quantities, in particular $b(z)$
and the amplitude $\omega(10'')$ are the same as in
Figure~\ref{fawz_err}.  Note again that the method we use to calculate
$\overline{\omega}_{\theta_{\rm max}}$ is likely to overestimate its
real value and therefore finite volume effects.

Finite volume errors become smaller due to the larger area covered,
while discreteness effects are reduced due to the larger number of
objects in the survey. Except at small $z$, where the effect of the
bias can make $E_1$ dominant again, the sizes of $E_1$ and $E_2$ are
of same order, and $E_3$, which contains the pure Poisson noise, is
now negligible.  Thus, a survey made with the Advanced Camera will no
 longer be dominated by shot-noise. In terms of sampling strategy, we
find that this kind of survey will be a good compromise between finite
volume effects and discreteness effects (e.g. Colombi, Szapudi \&
Szalay 1998), with a gain of more than a factor two for the total
cosmic errors compared to the present data, at least at the scale
considered here.
%
%--------------------------------------------------------
%--------------------------------------------------------
% 

\section{Conclusions}

In this paper, we have described the measurement of the galaxy
clustering as a function of $z$ in the HDF-South based on a
combination of optical HST data and VLT/ISAAC infrared data.  The
main results can be summarized as follows:

\begin{itemize}
\item The redshift distribution obtained for the HDF-South up to
 $I_{AB}\sim 27.5$ is consistent with that observed for the HDF-North.
 The peak of $N(z)$ is close to $z\sim 0.8$ with a decrease at
 $z\simeq 1$, a plateau from $1\le z \le 3$, followed by a decline of
 the number of objects up to $z\simeq 4.5$.
\item We have described an alternative approach to include
 photometric redshift uncertainties in the ACF measurements.  The
 method is based on a weighted measurement of the ACF taking into
 account the redshift probability distribution of each object.  This
 method makes it possible to extract the clustering signal with higher
 significance.  It will be interesting to implement and to test the
 method also for ground-based data which typically have larger
 photometric errors than HST data, and for less reliable photometric
 redshift, i.e. with a larger spread of the redshift probability
 distribution.  This approach can be extended to any kind of
 evolutionary studies based on photometric redshifts, like, for
 example, the luminosity function. 
\item We have compared the results of the clustering evolution
 obtained in the HDF-North and HDF-South.  Both are fully consistent
 within the Poissonian uncertainties. The new observations confirm our
 previous findings for the HDF-North (A99). The clustering amplitude
 shows a decrease between $0\le z \le 1$ and an increase at $z\ge
 1.5$.  The redshift range $1\le z \le 2$ seems to be a critical epoch
 where the HDF-galaxy clustering reaches a constant regime still
 difficult to characterize, due to the smallness of the present sample
 and to the critical redshift range for photometric redshift
 determination. Larger samples with the HST Advanced Camera will
 improve significantly the present picture.  The comparison with the
 behaviour of the underlying dark matter shows that the HDF-galaxy
 population is a nearly unbiased or anti-biased tracer of the dark
 matter distribution at $z\le 1$ and $z\le 1.5$ in $SCDM$ and $\Lambda
 CDM$ models, respectively. At higher redshift the clustering
 amplitude increases and the bias of this population too.  At $\langle
 z \rangle \simeq 3$, the bias is $b\simeq 3$ and $b\simeq 2$ for
 $SCDM$ and $\Lambda CDM$ models, respectively. This is in good
 agreement with the results we obtained for HDF-North (see also
 Magliocchetti \& Maddox 1999).  The typical minimum masses of the
 hosting dark matter haloes required to reproduce the observations in
 $SCDM$ model are $M_{\rm min}=10^{10} h^{-1} M_{\odot}$ at $z\le 1.5$
 and $M_{\rm min}\simeq 10^{11} h^{-1} M_{\odot}$ for $1.5 \le z \le
 3.5$ ($M_{\rm min}\le10^{10} h^{-1} M_{\odot}$ at $z\le 1.5$ and
 $M_{\rm min} \simeq 10^{11-11.5} h^{-1} M_{\odot}$ for $1.5 \le z \le
 3.5$ in $\Lambda CDM$ model).  At $ \langle z \rangle \simeq 4$, the
 clustering signal detected in the HDF-South is considerably smaller
 than the corresponding amplitude observed in the northern field, but,
 due to the very small sample (and, as a consequence, a large Poisson
 noise), the two results are still consistent within $1
 \sigma$. Again, larger samples are required at such a redshift.
\item In all our analysis we used errorbars assuming nearly Poisson
 statistics ($w(\theta) \ll 1$), as given by Landy \& Szalay
 (1993). To check {\em a posteriori} that such a procedure is valid,
 we used the analytical approach of Bernstein (1994), which fully
 describes the global budget of cosmic errors. In particular, the
 formulae obtained by Bernstein (1994) do not assume $\omega(\theta)
 \ll 1$ and take into account effects of higher-order statistics.  We
 checked that we recover the nearly-Poissonian contribution in
 Bernstein's calculations, and we found that it is indeed dominant in
 most regimes, except at small redshifts and at large angular scales,
 where the finite volume error (often called {\em cosmic variance})
 can become significant. Note that Bernstein's calculations neglect
 the edge effects, which can contribute to the errors (e.g. Szapudi \&
 Colombi 1996). However, by construction, the Landy \& Szalay
 estimator, that we used in our analysis, should minimize them to a
 large extent.

As a general conclusion of this paper, the HDF samples allowed us to
obtain a global picture of the redshift evolution of the galaxy
clustering, but with errorbars dominated by Poisson noise. Future
instruments, like the Advanced Camera, will improve the accuracy of
the measurement of $\omega(\theta)$ by at least a factor two, mainly
by reducing discreteness errors.  In particular, pure Poisson noise
will become subdominant and it will no longer be possible to neglect finite
volume effects in the analyses.
\end{itemize}

\section*{Acknowledgments.} 
We are grateful to Christophe Benoist, Narciso Benitez and Luiz da Costa
for general useful discussions. This work has been partially supported
by Italian MURST, CNR and ASI and by the TMR European network ``The
Formation and Evolution of Galaxies" under contract
n. ERBFMRX-CT96-086.

\end{document}